# A new Loan – Stock Financial Instrument

Alexander Morozovsky[1,2]
Bridge, 57/58 Floors, 2 World Trade Center,
New York, NY 10048
E-mail: alex@nyc.bridge.com
Phone: (212) 390-6126
Fax: (212) 390-6498

Rajan Narasimhan,
Overture Computing Corp. Ste 203
Jersey City, NJ 07302
E-mail: rajan@overture-computing.com
Phone: (201) 332 0700

Yuri Kholodenko,
Department of Chemistry,
University of Pennsylvania,
Philadelphia, PA 19104 - 6323
E-mail: kholoden@sas.upenn.edu
Phone: (215) 898 - 7149

**Abstract.**

A new financial instrument (a new kind of a loan) is introduced. The loan-stock instrument (LSI) combines fixed rate instruments (loans, etc.) with other financial instruments that have higher volatilities and returns (stocks, mutual funds, currencies, derivatives, options, etc.). This new loan depends on the value of underlying security (for example, stock) in such a way that when underlying security increases, the value of loan decreases and backwards. The procedure to create a risk free portfolio and a technique to fairly price the LSI is described. The philosophy behind this procedure is quite similar to the Black-Scholes formalism in option theory. Creation of the risk free portfolio is possible because the change in the underlying security offsets the change in the value of the loan (or the amount that the borrower has to repay). The new financial instrument takes an advantage of the fact that on average the stock market grows in time. It is beneficial for both the borrower and the lender. The LSI is more attractive for the borrower than the traditional loan is due to the decrease in the amount that has to be repaid. This attractiveness constitutes the benefit for the lender in terms of the market share among the borrowers. In addition, the lender can charge the extra premium.

---

[1] The ideas expressed in this article are the author's only and do not necessarily correspond to the views of Bridge, Overture Computing Corp., or University of Pennsylvania.
[2] Some material discussed in this article is protected by a pending patent, Serial number: 60/178,940, Filing date: 02-01-2000.

# Introduction.

There are many different financial instruments today that are used for different purposes. Stocks and bonds allow one to invest one's capital and get a return on one's investment. Loans allow a person to borrow capital for a cost (interest).

The idea behind a bond is that it is a way for an investor to earn an interest on an investment for a certain period of time. The idea behind loans is opposite to that of bonds. Loans allow a person to borrow money for a fixed period of time. After/during that period of time the borrower must return the entire borrowed amount (plus accumulated interest).

Many calculations in the financial industry are done with, what is known as, the risk free interest rates. For example, the price of treasury bills or bonds is calculated on the base of risk free interest rates.

In stocks, unlike bonds, an investor gets a bigger return. These returns come in the form of dividends or appreciation in the value of the stock. Higher returns however come with a price: investment in stocks is characterized by higher risk

# Overview of the idea.

In this communication we suggest a new financial instrument that leverages the higher returns possible with stocks, or stock-like instruments, to create a loan-like instrument that is beneficial for both lenders and borrowers. The lenders benefit from higher returns that are not accompanied by increased risk. The borrowers enjoy a lower cost of the loan that, however, might be associated with increased risk.

We conceive the simplest form of the new financial instrument as follows. A lender (for example a bank) will give a borrower a certain amount in cash. At the same time the lender will also buy a certain amount of stock or stock-like security (for example a mutual fund). When the value of the underlying security increases the amount that the borrower has to repay correspondingly (see below) decreases compared to the amount to be repaid when a traditional loan is employed. And vice versa, when the value of the underlying security decreases the amount that the borrower has to repay correspondingly increases. The borrower will make periodic payments that will depend on the loan amount and the value of the underlying stock. On average the stock market always seems to go up. Therefore, according to the described idea, the borrower will have to repay less than he/she would in a traditional loan situation. Thus the financial benefit of this new loan strategy for the borrower is obvious (at least on average). The financial benefit to the lender could come from an additional premium that the lender can charge from the borrower. The borrower's willingness to pay such an additional premium is due to the reduced repayment amount as described above. At the same time as we show below, the lender's market risk (risk of losing capital) could still be minimal and independent of the underlying security value. This is achieved by constructing a risk free portfolio that consists of the loan and the proper amount of the underlying security. It is important to understand that financial benefits of both the lender and the borrower come from an additional investment in a stock market that is associated with a particular loan and from the lender's ability to construct the risk free portfolio by properly balancing the relative amounts of the loan and the underling security. The situation is quite similar to the situation that arises when a risk free portfolio (that consists of an option and an underlying security) is created in option theory. We notice that financial instruments that could be used as the underlying security include, but are not limited to, stocks, generic mutual funds, mutual funds based on stock indices such as DJIA, currencies, different kind of derivatives, like futures, forwards and different kind of options, etc. The pricing model outlined below is equally applicable to all these cases.

The new LSI can be viewed as the instrument in which the lender invests in the stock market on behalf of the borrower. If so, one can ask, why wouldn't the borrower invest for himself? The answer is very simple. The borrower does not have money to do so. To have money invested for him, the borrower has to pay to the lender in the form of an additional premium to the lender. We will return to the question of this additional premium below.

This new loan-stock instrument (LSI) must have the following properties in order to exist and be marketable:

1. The amount of debt should decrease with time faster than it would when a traditional loan is employed. If not, the customers will not be interested in the LSI.
2. The value of the LSI should depend on the value of the underlying security.
3. One should be able to create a risk-neutral portfolio on the basis of this LSI in order to price it (introduced security).

In the most common model (absence of transaction costs, frictionless market, and so on) [1,2], the stock dynamic consists of two parts:

$$dS = \mu S dt + \sigma S dz \qquad (1)$$

Here S is the asset price. The first term describes predictable or deterministic return with μ being a measure of the average rate of growth of the asset price and dt is a small time interval. The second contribution (σSdz) reflects the random change in the asset price with volatility σ and dz is a Wiener process. We intend to construct now the risk-neutral portfolio in such a way that it will not be affected by the stock value changes.

The portfolio will consists of the loan P and a number of -Δ of the underlying security:

$$L = P - \Delta S \qquad (2)$$

The number Δ has yet to be determined. This step is quite similar to construction of the portfolio from an option and some amount of the underlying security in option theory. Following the usual derivation of the Black – Scholes formula [1,2] one immediately could obtain:

$$\frac{\partial P}{\partial t} + rS \frac{\partial P}{\partial S} + \frac{1}{2} \sigma^2 S^2 \frac{\partial^2 P}{\partial S^2} = rP \qquad (3)$$

where *r* is the risk free interest rate. *S* represents an underlying security with a standard deviation $\sigma$. Here *P* is a function of *S* and *t*. Black and Scholes solved this equation for European put and call options. However this equation could be applied for any other security.

### The solution for calculation of Loan-Stock instrument.

The value of the LSI introduced above depends on the value of the underlying security. As we have already mentioned the amount to be repaid by the borrower decreases when the value of the underlying security increases and vice versa. Different functional forms of such dependence that would satisfy the Black – Scholes equation (3) can be considered. For example, the value of the loan can be expressed as:

$$P(S, t) = A q(t) S^{-\beta} \qquad (4)$$

Here A is a constant that represents the value of the loan when $S = 1$. The term q(t) is calculated assuming arbitrage-free market conditions. The parameter $\beta$ (the loan-stock correlation parameter) determines how sensitive the loan is to the changes of the underlying security value (SRI). For example, if $\beta = 0.05$, changes in *S* will affect *P* less than they would when $\beta = 0.5$. In the limit $\beta = 0$ the LSI reduces to a traditional loan that is not coupled to any underlying security.

One way to calculate *q(t)* would be to assume that the underlying security could be modeled by a Wiener process. Substitution of P(S,t) from (4) into the Black – Scholes equation (3) immediately leads to the differential equation for q*(t)*. The solution of the equation is straightforward and the result is:

$$q(t) = e^{(r - \frac{1}{2}\sigma^2 \beta(\beta+1) + r\beta)t} q(t_0) \qquad (5)$$

Note that in the limit of $\beta = 0$ the increase in the loan value follows $e^{rt}$, as it is expected for the traditional loan. For the newly introduced LSI however the correlation parameter $\beta$ has to be chosen in accordance with the desired strength of the connection between the loan value and the value of the underlying security. For example, the lender might want to have the value of the loan to decrease 10 % in response to a 100% increase in the value of the underlying security. It directly follows from (4) that the value of $\beta = \log_2(1/0.9) = 0.152$ would satisfy this condition. A prime concern of the borrower, however, is how fast (on average) the loan value would decrease. Under the circumstances, the answer of course depends on the parameters of the underlying security dynamics. As we have noticed above (see (1)), the deterministic part of the return for the asset S that has an expected return $\mu$ is described by $\mu dt$. Therefore, the average growth of *S* over time follows the simple rule:

$$S(t) = S_0 e^{\mu t} \qquad (6)$$

It follows that on average it would take as long as τ = ln2/µ to have S increased 100% (that is twice).

In complete analogy with the option theory it follows from the above consideration that the amount of the underlying security (Δ) necessary to create a risk free portfolio is given by the rate of change of the loan value in respect to S:

$$\Delta = \frac{\partial P}{\partial S}, \tag{7}$$

Taking into account the connection between P(S,t) and S (formula (4)) this yields $\Delta = -\frac{\beta}{S} P(S,t)$. The additional investment required to buy Δ shares of the underlying security amounts to |Δ| S = β P. We notice that in all the formulas above the underlying asset price S can, as always, be modeled by the Poisson distribution or any other stochastic model.

We also notice that (4) is not the unique solution to the Black-Scholes equation (3). Other functional forms of the loan value dependence on the underlying security value are possible. These other loan-security combinations will also behave as a new financial instrument (LSI) described above. One additional example of the possible correlation between the values of the loan and the underlying security is given by:

$$P(S, t) = A(t)S^m - B(t)S^n \tag{8}$$

Any such solution (including (4) and (8)) can be used to fairly price the LSI. However, the first suggested solution (4) is the simplest solution for the Black-Scholes equation (3) that describes the proposed instrument.

## **A description of new security and relationship between suggested and existing instruments.**

Above we have introduced and described the new loan - stock financial instrument (LSI). Contrary to the traditional loan, the value of the LSI is coupled to (and therefore depends on) the value of an underlying stock (SRI). A borrower borrows from a lender a given loan amount, based on the terms and conditions of the LSI. The borrower then makes periodic payments based on the value of the underlying SRI and the amount borrowed. The lenders may choose to further insure themselves against possible default of the periodic payments or the loan amount by asking the borrowers to get an insurance policy, or getting one themselves. Given the loan amount L and a SRI of value S one can create the LSI using any solution to the equation (3) that acts like the new financial instrument just described.

The LSI is a kind of a negative security that behaves like a SRI but has a negative value. Table 1 shows a simplified relationship between loans, bonds (FRI), SRIs (stocks) and their possible applications.

Table 1:

|  | **Investment decisions made on the basis of interest rate** | **Investment decisions are made on the basis of expected return, volatility** |
|---|---|---|
| **Positive Security** | Treasury Bonds, Corporate Bonds, bill etc. FRI | Individual Stocks, Stock Funds, Options, etc SRI |
| **Negative Security** | Loans | New Financial Instrument LSI |

### **Illustrative Example.**

In this section we provide a reader with an example that illustrates the idea of the new loan instrument (LSI). The parameters of the problem are assumed to be as follows:

1) The borrower requires a loan amount of $100,000.
2) $\beta = 0.347$.
3) $\mu = 0.2$ (A historical expected return of the SRI is 20%).
4) $r = 0.1$ (Risk free interest rate is 10%).
5) $\sigma = 0.1$ (Historical Variance of the SRI).

For simplicity we also assume that the borrower does not make any periodic payments. The results of the calculations are shown in Table 2. As usual, the amount owed when traditional loan is employed follows $e^{rt}$. The amount owed with the LSI is calculated based on the formulas (4) and (5). In agreement with the conditions of the example the combination of normalization parameters (A, $q(t_0)$, $S_0$) is chosen to satisfy the initial condition for $P(S,t)$ to be $100,000 on the beginning date of the loan. The underlying security dynamic is assumed to be described by formula (6). (This of course is true only on average. In reality the stock value is a subject to stochastic fluctuations). Finally, the value of the stock part (column four in Table 2) is simply $S^{-\beta}$.

    It also shows how much SRI the bank needs to hold to break even, i.e. to make the same amount of money that the bank would make if it held a risk free FRI for an amount equal to the sum of the loan amount and the value of the SRI.

Table 2.

| Years | The amount owed with a traditional loan(FRI), $ | The amount owed with the (LSI),$ (Subject to statistical fluctuations). | Value of stock part |
|---|---|---|---|
| 0 | 100000 | 100000 | 1.00 |
| 1 | 110517.09 | 106498.74 | 0.93 |
| 2 | 122140.28 | 113419.81 | 0.87 |
| 3 | 134985.88 | 120790.67 | 0.81 |
| 4 | 149182.47 | 128640.54 | 0.76 |
| 5 | 164872.13 | 137000.55 | 0.71 |
| 6 | 182211.88 | 145903.86 | 0.66 |
| 7 | 201375.27 | 155385.77 | 0.62 |
| 8 | 222554.09 | 165483.89 | 0.57 |
| 9 | 245960.31 | 176238.25 | 0.54 |
| 10 | 271828.18 | 187691.51 | 0.50 |

Under the assumption of the continuous re-balancing of the number of shares, the amount of money that the bank has (without additional premium) will be exactly the same as the growth of a normal fixed rate loan. This is due to the fact that the change in the underlying security will offset the change in the value of the new loan security (or the amount that the borrower has to repay). This means that if the underlying security price increases, the value of the new loan-security will decrease in such a way that the sum of the two amounts will grow exponentially in time with the same rate as the fixed rate loan would. The opposite is also true. If the stock price decreases then the value of the new loan-security will increase in such a way that the sum will again grow in exactly the same way as the traditional fixed rate loan would.

However, a strategy for the bank to earn some *additional* return has to be specified. The lender proceeds as follows:

1. At the time when the loan is initiated the lender also buys some units of SRI's (for example, shares of stock). The number of shares is determined from the formula (7). With the parameters described above and assuming the price of a share of the underlying stock to be just $1.00 one gets $\Delta = 34700$ with an additional investment of $34,700.00.
2. The lender periodically changes the number of shares depending on the value of the SRI. For example, the lender might want to recalculate the number of shares held (and therefore buy or sell some shares of stock) at the end of a certain period of time (a year for example) based on the value of the underlying stock at that time.
3. Based on the financial benefits of the new instrument for the borrower (Table 2), the lender can afford to charge an additional premium from the borrower. Of course the amount of the premium could not exceed the financial benefit that the borrower obtains in the LSI comparing with the traditional loan. For example, for the parameters described above, the additional premium is associated with the additional investment of $34,700.00 at the beginning of the loan. We stress however that both the amount of the loan ($100,000) and the amount of the additional investment ($34,700.00) keep accumulating interest with the fixed rate (r = 0.1 in our example).

Therefore, in addition to the competitive attractiveness of the LSI for the borrower this additional premium constitutes the financial benefit for the lender (as compared to the traditional loan).

This consideration is illustrated in Table 3. The third column lists the number of shares to be bought by the lender to balance its portfolio. (We again assumed that the underlying security dynamics is described by (6). In reality this assumption is true only on average). Maximum possible premium for the lender (i.e. the financial benefit of the borrower that is equal to the difference between the amounts that have to be repaid by him in a traditional loan and in the LSI) is enlisted in the fourth column.

Table 3

| Years | Growth of the money owed to the bank(FRI) | Number of stock shares if price of one share = 1 at the beginning of the loan. | Additional possible maximal return for the bank |
|---|---|---|---|
| 0 | 100000 | 34700 | 0.00 |
| 1 | 110517.09 | 30256.25 | 4018.35 |
| 2 | 122140.28 | 26381.57 | 8720.46 |
| 3 | 134985.88 | 23003.09 | 14195.21 |
| 4 | 149182.47 | 20057.27 | 20541.93 |
| 5 | 164872.13 | 17488.69 | 27871.57 |
| 6 | 182211.88 | 15249.05 | 36308.02 |
| 7 | 201375.27 | 13296.23 | 45989.50 |
| 8 | 222554.09 | 11593.49 | 57070.21 |
| 9 | 245960.31 | 10108.80 | 69722.06 |
| 10 | 271828.18 | 8814.25 | 84136.67 |

## **Possible Disadvantages.**

Although the new financial instrument presents several advantages (described above) over the traditional loans widely used today; there are certain disadvantages associated with the LSI that should be mentioned.

Borrower's disadvantage: the borrower is now exposed to the market risk for the amount invested in the stocks.

Lender's disadvantage: default risk for the lender increases. If the stock value decreases, the possibility that the borrower may default increases. However since on average the stock market goes up, this risk decreases with time. The bank can further insure against default risk by charging a premium (as described above) or requiring the borrower to take an insurance policy against such an event.

## A new kind of bond created by combining the LSI and SRI.

As was shown above, the LSI grows in time with the fixed interest rate r. Therefore a new kind of bond can be created that combines the LSI (loan-like instrument) and SRI (stock or stock-like instrument). This new bond portfolio could be packaged and sold to interested 3$^{rd}$ parties.

## Conclusion.

We have described a new type of combined loan-stock financial instrument (LSI). The risk free portfolio is constructed by combining a traditional loan with some amount of underlying security. The philosophy behind the LSI is quite similar to the Black-Scholes formalism in option theory. As a result the mathematical formulations are as well similar. One could choose more favorable conditions for borrowers by increasing $\beta$, the parameter that regulates the connection between the loan and the underlying security in the combined LSI. The increased value of β implies that more money is invested in stocks or stock like securities on behalf of the borrowers by the lender. However this is also more risky for the borrower (and thus for the lender) since the borrower would find himself more exposed to the market volatility and would have to pay higher periodic premiums (monthly, bi-monthly, quarterly, etc.) in the event of the decrease of the underlying security value. The new financial instrument takes advantage of the fact that on average the stock market grows in time. The LSI is beneficial for both the borrower and the lender. The borrower benefits due to the decreased amount that has to be repaid. This instrument is therefore more attractive for the borrower than the traditional loan. This attractiveness constitutes the benefit for the lender in terms of market share among the borrowers. In addition, the lender can charge an extra premium. This premium could partially be used to cover the transaction costs and an insurance component. In addition, sellers of the LSI could also create new mortgage-backed-stock securities.